\documentclass[final,1p,times,authoryear]{elsarticle}
\usepackage{amssymb}
\usepackage{amsmath}
\usepackage{amsthm}

\usepackage{adjustbox}
\usepackage{float}

\journal{Internet of Things}

\begin{document}

\begin{frontmatter}

\author[aff4]{Mahadev Sunil Kumar\corref{cor1}}
\ead{mahadevsunilkumar03@gmail.com}
\ead[url]{https://www.mahadevsunilkumar.com}

\author[aff2]{Arnab Raha}
\ead{arnab.raha@intel.com}
\ead[url]{https://sites.google.com/site/arnaverse}

\author[aff3]{Debayan Das}
\ead{debayandas@iisc.ac.in}
\ead[url]{https://eecs.iisc.ac.in/people/debayan-das/}

\author[aff4]{Gopakumar G}
\ead{gopakumarg@am.amrita.edu}
\ead[url]{https://www.amrita.edu/faculty/gopakumarg/}

\author[aff5]{Rounak Chatterjee}
\ead{rounakchatterjee007@gmail.com} 

\author[aff6]{Amitava Mukherjee}
\ead{amitava@dubai.bits-pilani.ac.in}
\ead[url]{https://www.bits-pilani.ac.in/dubai/dr-amitava-mukherjee/}

\affiliation[aff4]{organization={Amrita Vishwa Vidyapeetham},
            city={Kollam},
            country={India}}

\affiliation[aff2]{organization={Intel Corporation},
            city={Santa Clara},
            country={United States of America}}

\affiliation[aff3]{organization={Indian Institute of Science},
            city={Bangalore},
            country={India}}

\affiliation[aff5]{organization={Analog Devices, Inc.},
            city={Edinburgh},
            country={United Kingdom}}

\affiliation[aff6]{organization={Birla Institute of Technology and Science-Pilani, Dubai Campus},
            city={Dubai},
            country={United Arab Emirates}}

\cortext[cor1]{Corresponding author}

\title{SlimEdge: Performance and Device Aware Distributed DNN Deployment on Resource-Constrained Edge Hardware} 

\begin{abstract}
Distributed deep neural networks (DNNs) have become central to modern computer vision, yet their deployment on resource-constrained edge devices remains hindered by substantial parameter counts, computational demands, and the probability of device failure. Here, we present an approach to the efficient deployment of distributed DNNs that jointly respect hardware limitations, preserve task performance, and remain robust to partial system failures. Our method integrates structured model pruning with a multi-objective optimization framework to tailor network capacity for heterogeneous device constraints, while explicitly accounting for device availability and failure probability during deployment. We demonstrate this framework using Multi-View Convolutional Neural Networks (MVCNN), a state-of-the-art architecture for 3D object recognition, by quantifying the contribution of individual views to classification accuracy and allocating pruning budgets accordingly. Experimental results show that the resulting models satisfy user-specified bounds on accuracy and memory footprint, even under multiple simultaneous device failures. The inference time is reduced by factors up to 4.7$\times$ across diverse simulated device configurations. These findings suggest that performance-aware, view-adaptive, and failure-resilient compression provides a viable pathway for deploying complex vision models in distributed edge environments.
\end{abstract}

%

\end{frontmatter}
%
%
\section{Introduction}
\label{introduction}
The deployment of DNNs on resource-constrained edge devices has become a major challenge in the widespread availability of Internet of Things (IoT) ecosystems. While modern architecture achieves high accuracy in tasks such as 3D object recognition, their computational and memory requirements often exceed the capabilities of low-power hardware commonly found in edge devices. This problem raises a fundamental question: how can complex vision models be adapted to heterogeneous edge devices without compromising task performance?

\begin{figure}[h]
	\centering
	\includegraphics[width=0.8\textwidth]{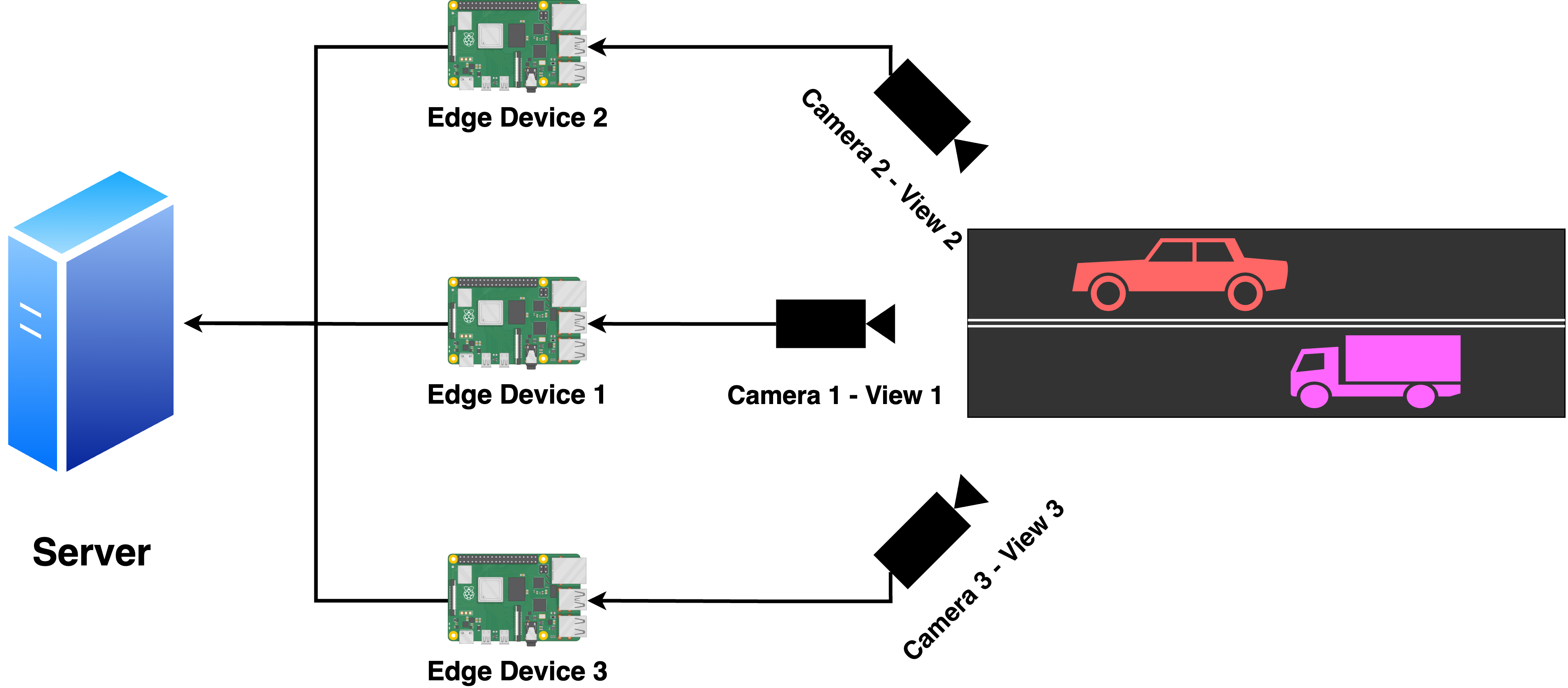}
	\caption{Roadside Vehicle Monitoring System using 3 Cameras}
	\label{fig:fig-1}
\end{figure}

Multi-View inference offers a natural setting in which this problem manifests. In Multi-View Convolutional Neural Network (MVCNN) \citep{su_multi-view_2015}, a state-of-the-art (SOTA) architecture for 3D object recognition, where distinct 2D images of an object are processed in parallel by separate feature extractors. These are typically deployed across multiple edge nodes, before feature aggregation, and classification at a central server. As illustrated in Figure \ref{fig:fig-1}, such a configuration might involve a roadside monitoring system, where each camera feeds into a local edge device tasked with extracting view-specific features for classifying vehicles such as cars and trucks to identify traffic violations and road traffic conditions. But, while this distributed design enables spatial parallelism, it also exposes the system to the combined constraints of per-device memory budgets, variable compute capacities, and global accuracy requirements. \\
Uniform model compression, such as applying identical pruning ratios across all views, fails to account for two critical asymmetries: (i) the unequal contribution of individual views to classification accuracy, and (ii) the heterogeneous computational capabilities of the devices hosting them. Aggressive pruning of highly informative views on capable hardware, or insufficient compression on devices with lower memory, can degrade the accuracy or violate memory constraints, respectively. Moreover, the failure of one of these devices would greatly degrade the performance of the framework, making it infeasible for real-world deployment.
While prior research has explored model compression and distributed inference independently, SlimEdge occupies a unique position by treating these as a unified system-level challenge. Unlike standard uniform pruning, which assumes all input views are equally valuable, our framework recognizes the informational asymmetry inherent in multi-view data. Furthermore, while existing hardware-aware methods focus on static performance, SlimEdge is one of the first to integrate dynamic device availability into the pruning optimization loop, ensuring the system remains functional even when the physical topology changes mid-inference. The technical advantages of SlimEdge over current state-of-the-art deployment strategies are summarized in Table \ref{tab:tab-1}.

\begin{table}[h]
	\centering
	\begin{adjustbox}{width=\textwidth}
		\begin{tabular}{|p{3.5cm}|p{3cm}|p{3cm}|p{3cm}|}
			\hline
			\textbf{Feature} & \textbf{Uniform Pruning} & \textbf{Hardware-Aware Pruning} & \textbf{SlimEdge} \\ 
			\hline
			\textbf{Allocation Metric} & Uniform across layers & Latency/FLOPs & View Importance + Latency \\ 
			\hline
			\textbf{Node Failure Resilience} & System Failure & Static/No Recovery & Dynamic Pruning Reallocation \\ 
			\hline
			\textbf{Heterogeneity Support} & None & Device Specific & Per-Device Capacity \& Device Performance \\ 
			\hline
			\textbf{Optimization Target} & Accuracy & Model Size + Accuracy & Model Size, Accuracy, Inference Time \\ 
			\hline
		\end{tabular}
	\end{adjustbox}
	\caption{Comparison of Deployment Frameworks for Multi-View Inference}
	\label{tab:tab-1}
\end{table}

To address these issues, we introduce SlimEdge, a framework for view and device-aware model compression. Our approach begins by quantifying the importance of each view, as well as the per-view device metrics to guide a pruning process that respects user-defined constraints on accuracy and model size. SlimEdge then generates a set of models, each tailored to its host device while collectively satisfying global performance targets. Our framework actively updates the architecture, with respect to the status of the devices involved.
The main contributions of the paper are:
\begin{enumerate}
    \item Metric Formulation: We propose a joint objective function that couples view-importance (via first-order Taylor expansion) with real-time hardware latency, moving beyond purely accuracy-centric or purely hardware-centric metrics.
    \item Optimization Strategy: We introduce a biased-initialization NSGA-II \citep{deb_fast_2002} approach. By using Beta-distribution sampling to seed the population with importance-aware candidates, we achieve faster convergence in high-dimensional pruning spaces compared to standard genetic algorithms.
	\item Failure Resilience: We implement a dynamic reconfiguration logic that automatically reallocates pruning budgets across active nodes upon device failure, a capability absent in current static distributed inference frameworks.
    \item Systematic Simulation: We evaluate the framework across 1,000 simulated heterogeneous configurations, demonstrating latency reduction up to 4.70$\times$ over uniform pruning baselines.
\end{enumerate}
This work transcends static, one-size-fits-all compression, adopting a dynamic deployment strategy. It considers the diverse computing and memory capabilities of each device by compressing each model individually based on the device's specifications. Consequently, the resulting models not only adhere to stringent memory constraints but also significantly reduce end-to-end latency by factors ranging from 1.2$\times$ to 4.70$\times$ across heterogeneous platforms. Moreover, these models maintain classification accuracy above user-defined thresholds while compensating for device failure, thus adding robustness and fault tolerance to the framework.

\section{Related work}
\label{related-work}
The deployment of Deep Neural Networks (DNNs) on resource-constrained edge hardware has become a pivotal challenge at the intersection of machine learning and embedded systems. While contemporary architectures excel in vision tasks, their computational and memory requirements frequently surpass the capabilities of low-power platforms like microcontrollers or single-board computers. This section organizes the existing literature into key thematic areas to highlight the current gaps in distributed edge intelligence.
\subsection{Model Compression for Edge Intelligence}
The disparity between model size and hardware capacity has spurred research on model compression, encompassing techniques such as quantization, coding, and pruning. These methods aim to reduce the parameter count and arithmetic complexity of models while striving to maintain their representational capacity \citep{liang_pruning_2021}. Among these, pruning has gained particular attention due to its compatibility with existing hardware and software stacks. Early methods involved removing individual weights, while more recent techniques operate at a structural level, eliminating entire neurons, channels, or filters \citep{molchanov_pruning_2017}. Filter pruning reduces both memory usage and floating-point operations without requiring specialized inference engines. 
While effective for individual models, most pruning strategies assume a homogeneous deployment target and apply uniform compression. They lack the mechanism to account for the inherent heterogeneity and informational asymmetry present in distributed multi-system environments.
\subsection{Distributed and Multi-View Inference}
Efforts to distribute DNN inference have explored layer-wise partitioning \citep{zhao_deepthings_2018} or input-dependent offloading \citep{chen_searching_2018}. These methods enhance latency through cloud-edge collaboration but often treat the model as a monolithic entity, failing to consider the unique physical constraints of individual nodes during compression. This limitation is particularly evident in multi-view settings, such as the MVCNN architecture \citep{su_multi-view_2015}, which achieves state-of-the-art performance in 3D shape recognition by fusing features from multiple 2D projections. Subsequent refinements, including Latent-MVCNN \citep{yu_latent-mvcnn_2020} and accuracy-optimized variants \citep{angrish_mvcnn_2021}, have further demonstrated the effectiveness of view-based fusion. Existing works do not address how to adapt per-view feature extractors to heterogeneous edge devices with varying memory budgets and compute capabilities.
\subsection{Hardware-Aware and System-Level Optimization}
Recent research has explored hardware-aware neural architecture search (NAS) \citep{baker_designing_2017} to bridge the gap between model design and deployment. Others have implemented coarse-grained pruning on FPGA-based platforms \citep{yao_efficient_2019}. However, NAS methods remain computationally expensive and difficult to integrate into pre-trained models, while FPGA-based approaches lack the fine-grained adaptability required for heterogeneous, resource-constrained edge environments. A significant limitation remains: existing approaches separate structural importance from deployment context. They fail to recognize that the contribution of each view to final classification accuracy is not consistent. Consequently, compression is rarely tailored to the specific performance of each device. Recent work has also explored system-level approximations for multi-view deep neural networks, proposing deployment-aware simplifications to reduce computational overhead in distributed settings \citep{das_toward_2024}. This approach performs approximation at the system level rather than adapting pruning intensity per view or per device. While such techniques effectively reduce global computational cost, they do not explicitly integrate per-device memory heterogeneity, dynamic availability, or view-specific informational importance into the optimization loop. The pruning framework proposed in this work can complement these system-level approximations by serving as a compute-aware compression layer, enabling fine-grained per-view adaptation within heterogeneous edge deployments.
\subsection{Fault Tolerance in Edge ML}
Real-world edge deployments are characterized by hardware volatility and intermittent node failure. Current model compression and distribution strategies largely ignore these dynamics. Without a recovery or re-optimization mechanism, the failure of a single device can significantly degrade the performance of the entire distributed framework, making it infeasible for mission-critical applications.

\section{Methodology}
\label{methodology}
\begin{figure}[ht]
	\centering
	\includegraphics[width=0.7\textwidth]{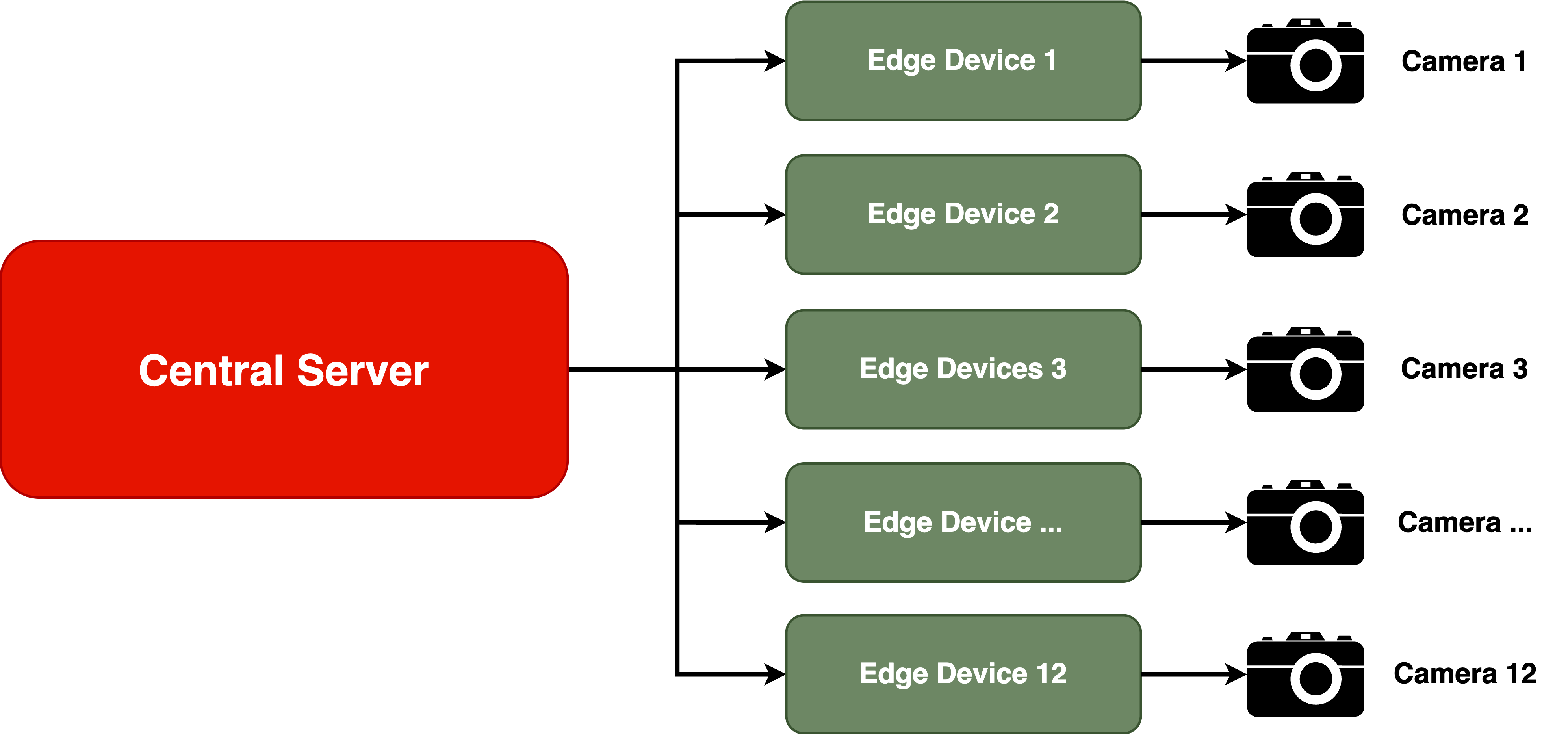}
	\caption{Distributed Multi-View Inference System Under Consideration}\label{fig:fig-2}
\end{figure}

SlimEdge addresses the deployment of MVCNNs across heterogeneous edge devices under hard constraints on model size, accuracy, and inference latency. The framework is orchestrated by a central server (Figure \ref{fig:fig-2}) and integrates structured pruning, view importance quantification, multi-objective optimization, and fault tolerance within a unified deployment framework (Figure \ref{fig:fig-4}). The server generates deployment-aware, per-device optimized models that satisfy user-defined accuracy requirements, device-specific memory and compute constraints, and partial device failure conditions. The mathematical symbols used throughout the methodology are summarized in Table \ref{tab:tab-2}.
\subsection{ModelNet40 dataset}
\begin{figure}[ht]
	\centering
	\includegraphics[width=0.9\textwidth]{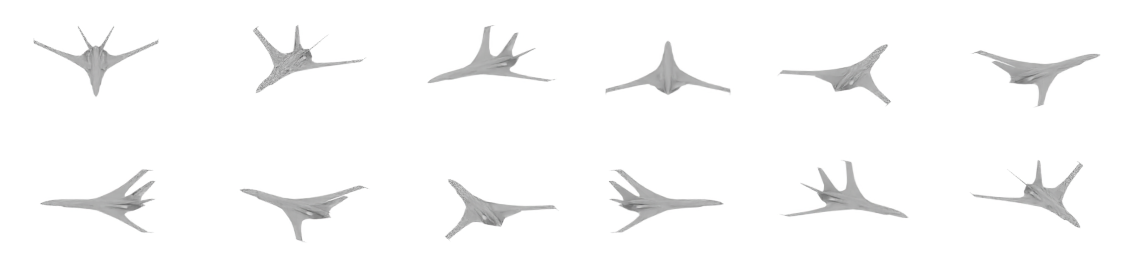}
	\caption{Views of "airplane\_0006" from 12 different views from the ModelNet40 dataset}\label{fig:fig-3}
\end{figure}

The ModelNet40 dataset \citep{wu_3d_2015} serves as a standard benchmark for 3D object recognition, comprising 12,311 CAD models categorized into 40 semantic classes. This dataset allows the use of 3D objects to generate 2D views, which can then be utilized to simulate a distributed environment with multiple edge devices capturing diverse perspectives. Out of the 12,311 CAD models, 9,843 are designated for training, while 2,468 are reserved for testing (80\% for training, 20\% for testing). To replicate real-world multi-view sensing, each 3D object is rendered from 12 virtual positions, uniformly spaced at 30° azimuthal intervals and fixed at a 30° elevation angle \citep{su_multi-view_2015}. Figure \ref{fig:fig-3} illustrates the 12 views when the 3D object “airplane\_0006” is employed. This results in a consistent set of 12 two-dimensional projections for each object, which serve as inputs to the distributed feature extractors. The 12 distinct views depicted in Figure \ref{fig:fig-3} enable the model to learn effectively, ultimately leading to improved classification performance. 
\begin{table}[ht]
	\centering
	\begin{adjustbox}{width=\textwidth}
		\begin{tabular}{|p{1.5cm}|p{5cm}|p{5.5cm}|}
			\hline
			\textbf{Symbol} & \textbf{Definition} & \textbf{Context} \\ 
			\hline
			$V; v \in V$ & Number of Views/Devices & Fixed at 12 for MVCNN \\ 
			\hline
			$p_v$ & Pruning ratio for view $v$ & Range: $[0,1]$, 1.0 = offline \\ 
			\hline
			$I_v$ & Normalized View Importance & Based on LightGBM feature attribution \\ 
			\hline
			$D_v$ & Device Performance Factor & Higher values indicate faster devices \\ 
			\hline
			$M_v$ & Device Memory Budget & Hard constraint for model fitting \\ 
			\hline
			$\mathcal{A}_{\min}$ & Minimum Accuracy Threshold & User-defined constraint \\ 
			\hline
		\end{tabular}
	\end{adjustbox}
	\caption{Summary of Mathematical Notation Used}
	\label{tab:tab-2}
\end{table}

\subsection{SlimEdge architecture}
We adopt the Multi-View Convolutional Neural Network \citep{su_multi-view_2015} paradigm, adapting it for distributed edge deployment. The architecture comprises two distinct functional stages: a distributed feature extractor and a centralized aggregation stage. \\
Distributed Feature Extractor: This stage involves $V=12$ independent convolutional subnetworks, $\phi_v$, deployed across separate edge nodes. We utilize a modified VGG11 backbone, pre-trained on ImageNet1K \citep{deng_imagenet_2009}. Each edge device $v$ processes a local view  $x_v$ to produce a lower-dimensional feature embedding $f_v$:  $f_v=\phi_v\left(x_v\right), \quad v=1, \dots ,V$.\\
This decoupled design enables SlimEdge to independently extract spatial information from each view, thereby facilitating parallel inference.
\begin{figure}[ht]
	\centering
	\includegraphics[width=1.0\textwidth]{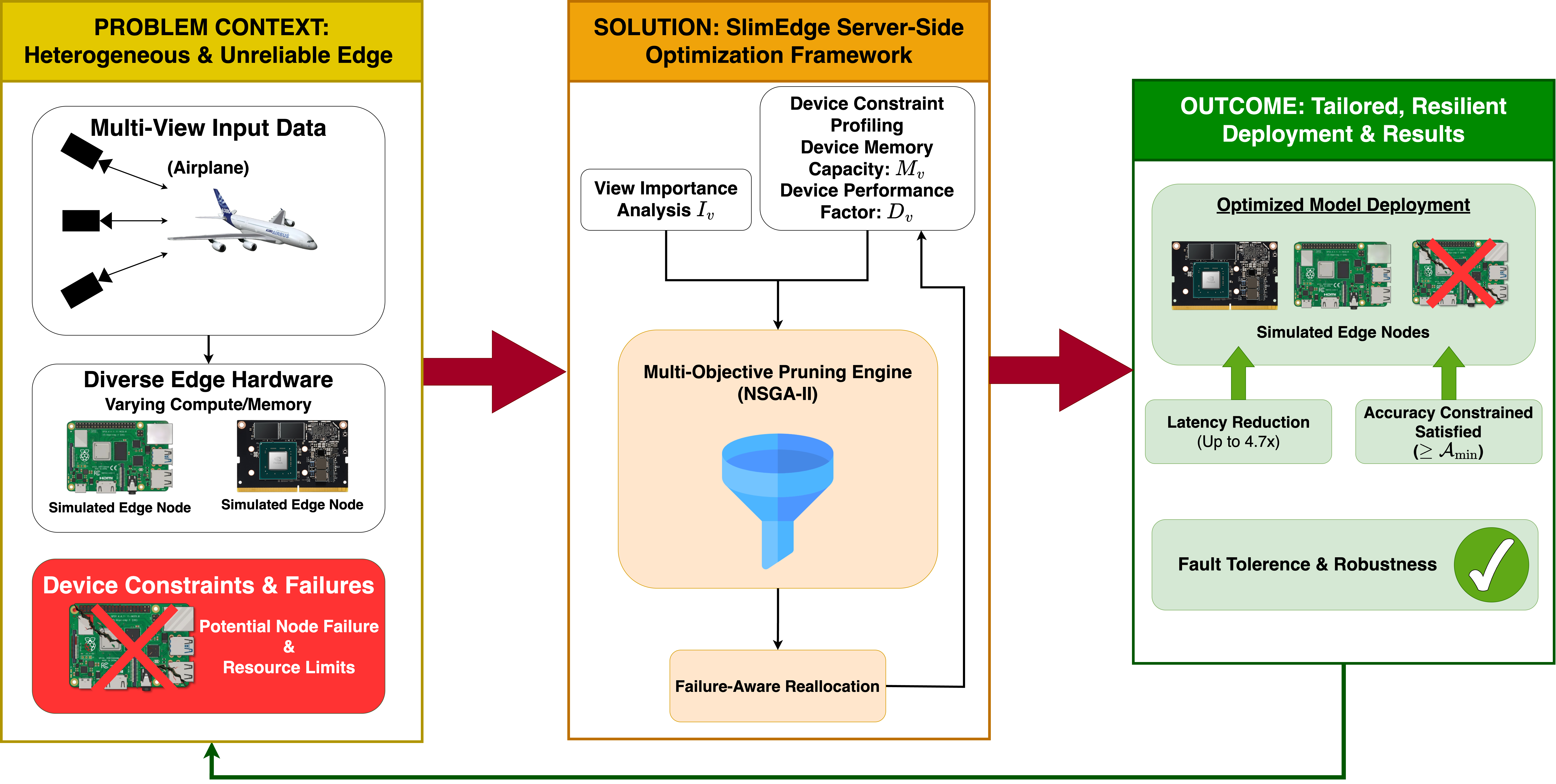}
	\caption{Overview of the SlimEdge framework. Multi-view inputs and heterogeneous device constraints are processed by a server-side multi-objective pruning engine (NSGA-II), which generates deployment-aware configurations satisfying latency, accuracy, memory, and failure constraints.}\label{fig:fig-4}
\end{figure}
View Pooling and Classification: After extraction, the feature embeddings, $f_1,\ldots f_V$ are transmitted to a central server, while a view-pooling layer aggregates these inputs into a single global descriptor $F$ using element-wise maximum-pooling: $F = \max_{v=1,\dots,V} f_v$
This aggregation mechanism guarantees permutation invariance, making the model resilient to alterations in the sequence of camera inputs. $F$ is subsequently passed to a central classifier, a fully connected network that predicts the final object class $y$.

\subsection{SlimEdge}
Figure \ref{fig:fig-4} illustrates the end-to-end SlimEdge workflow, from multi-view inputs and device profiling to server-side optimization and deployment. The application-specific requirements, i.e., the base model, and the desired accuracy for the targeted application, and the device specific constraints are sent to the server. The device constraints include the memory capacity and computing capacity of the device, and the views captured by it. The server runs the optimization process by evaluating view importance and integrating device-specific memory and compute constraints into a unified deployment-aware pruning strategy. The optimized model for each edge device is then created by the server and deployed on the respective edge devices. Therefore, the server performs two tasks: running the optimization framework, SlimEdge, and pooling the features received from the edge devices for final classification. In our experiments, we have selected distributed object recognition as the targeted application and have chosen the Multi-View Convolutional Neural Network (MVCNN) \citep{su_multi-view_2015} architecture as the base model.
A detailed explanation of the SlimEdge is provided in section \ref{subsubsec:components-of-slimedge}.

\subsubsection{Components of SlimEdge}
\label{subsubsec:components-of-slimedge}
\begin{figure}[ht]
	\centering
	\includegraphics[width=1.0\textwidth]{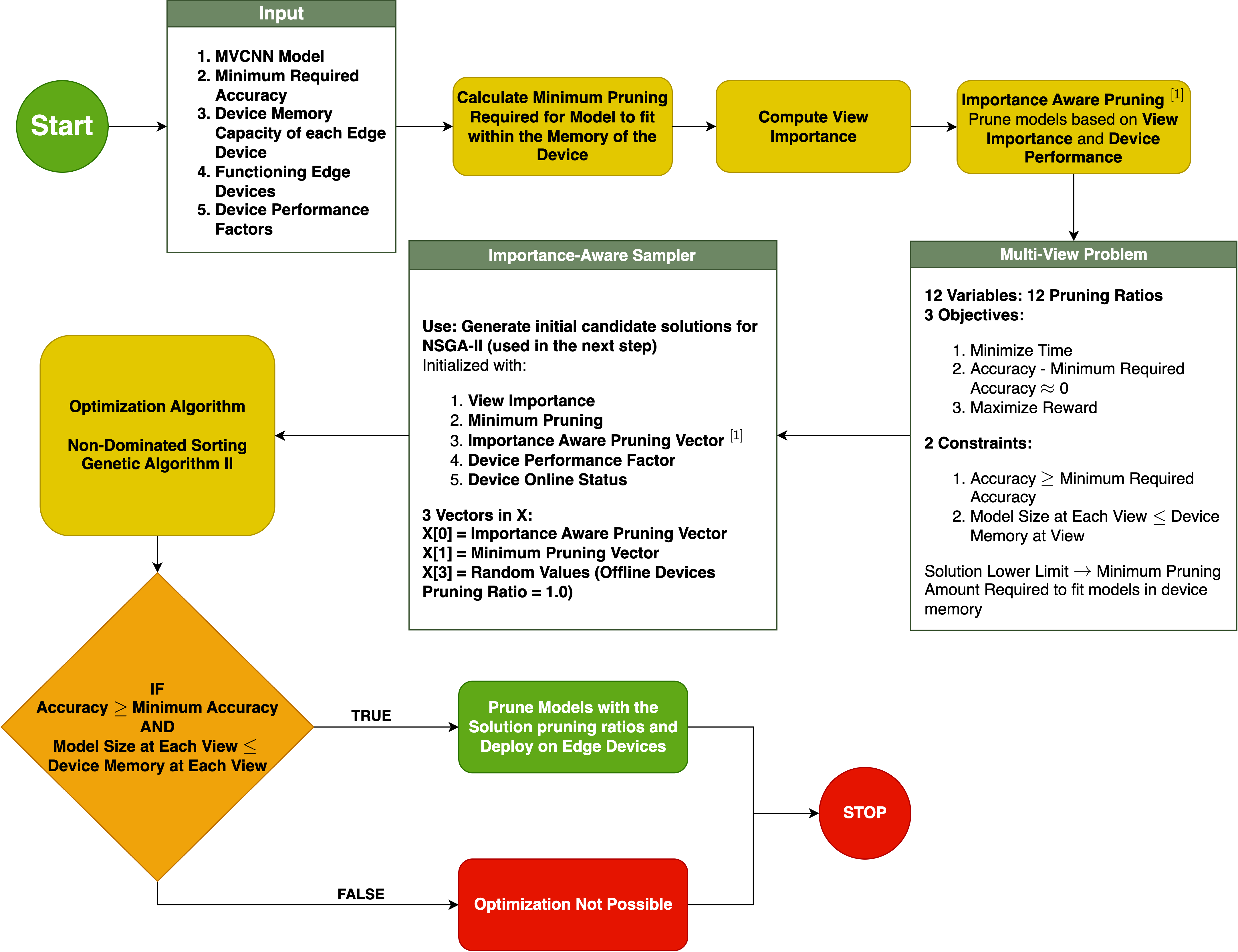}
	\caption{Component-Level Flowchart of SlimEdge}\label{fig:fig-5}
\end{figure}

SlimEdge, as depicted in Figure \ref{fig:fig-5}, is orchestrated by a central server. Unlike treating all views uniformly, the framework dynamically adjusts pruning intensity for each view based on its informational importance and the capabilities of the device. This approach effectively balances the global accuracy requirements with the resource constraints of individual devices.
\subsubsection{Model pruning}
To address memory and latency constraints, filter-level pruning was applied to each view-specific feature extractor. This component runs on the central server, and this approach allowed SlimEdge to reduce the model size while maintaining its performance. Following \cite{molchanov_pruning_2017}, the importance of the n-th filter was estimated using a first-order Taylor expansion of the loss function as shown in Equation \ref{eq:eq-1}.
\begin{equation}
	\mathcal{I}_j
    =
    \left|
    \frac{1}{B}
    \sum_{b=1}^{B}
    \left\langle
    \frac{\partial \mathcal{L}}{\partial f_j^{(b)}},
    f_j^{(b)}
    \right\rangle
    \right|
	\label{eq:eq-1}
\end{equation}
Here, $f_j^{(b)}$ denotes the feature map produced by the $j$-th filter for the $b$-th sample in a mini-batch of size B, $\mathcal{L}$ is the classification loss, and $\langle \cdot , \cdot \rangle$ denotes the Frobenius inner product over all spatial locations of the feature map. \\
Filters with the lowest importance scores were iteratively removed, rather than in a single pass, to prevent a sudden decline in representational capacity. The pruning schedule followed an exponential curve, with the pruning fraction at step $i$, $p^{\text{step}}_i$ calculated as in Equation \ref{eq:eq-2}.
\begin{equation}
	p^{\text{step}}_i=\frac{e^{-ki}}{\sum_{j=0}^{T}e^{-kj}}\cdot P
	\label{eq:eq-2}
\end{equation}
Where $P$ is the total pruning fraction, $k$ the decay rate, and $T$ the number of steps (Equation \ref{eq:eq-2}). After each pruning step, the model was fine-tuned to regain lost accuracy. This approach is more efficient in preserving performance compared to one-shot compression as explained by the Lottery Ticket Hypothesis \citep{frankle_lottery_2019}.

\subsubsection{Calculating minimum pruning vector}
\label{sub:min-pruning-vec}
To accelerate convergence, the minimum pruning required to satisfy per-device memory limits was computed using the Newton-Raphson method \citep{raphson_analysis_1697}. Here, $p_v$ is the ratio of filters pruned to the number of filters present in the original model. Given a target model size, the necessary pruning ratio $p_v$ was determined iteratively using the Newton-Raphson method. For each view v, let $p_v\in\left[0,1\right]$ be the pruning ratio (0 = unpruned, larger = more pruning) and $M_v$ the device memory budget at the view. The minimum pruning for device $v$  is defined as the smallest pruning that satisfies the memory constraint as in Equation \ref{eq:eq-3}:
\begin{equation}
	p_v^{\min} = \min \left\{ p_v \in [0,1] : \hat{S}(p_v) \le M_v \right\}
	\label{eq:eq-3}
\end{equation}
Implemented by solving the boundary equation $\hat{S}\left(p_v\right)=M_v$ via the Newton-Raphson method as in Equation \ref{eq:eq-4} where $\hat{S}\left(p\right)$ is the size of the model at pruning ratio $p$.
\begin{equation}
	p^{\left(t+1\right)}=p^{\left(t\right)}-\frac{f\left(p^{\left(t\right)}\right)}{f^\prime\left(p^{\left(t\right)}\right)}, \quad f\left(p_v\right)=\hat{S}\left(p_v\right)-M_v
	\label{eq:eq-4}
\end{equation}
The iterate is clipped to $\left[0,1\right]$. If $M_v<\hat{S}\left(1.0\right)$ (budget smaller than the surrogate’s achievable minimum), we return $p_v^{\min}=1.0$. Offline devices are fixed to $p_v^{\min}=1.0$. Finally, the minimum pruning vector $\mathbf{p^{\min}}=\left[p_1^{\min},p_2^{\min},\ldots,p_{12}^{\min}\right]$.

\subsubsection{Computing view importance}
\label{sub:comp-view-imp}
To quantify the importance of each view, we constructed a supervised regression vector space that maps a device/view availability and per-view pruning configuration to the resulting multi-view classification performance. Each row in the dataset, which had 93000 rows, contains 12 view-specific entries and measured accuracy.  Each column from $v_1,\ldots,v_{12}$ encodes the pruning ratio applied to each view. $v_i\in\left[0,1\right]$ denotes the pruning level used for view $i$. The last column contains the accuracy of the configuration. To achieve this, we created a matrix of configurations of shape 93000 x 12 where each element is a random pruning ratio. For each view, pruning values were sampled using a Latin-hypercube-like \citep{mckay_comparison_1979} strategy to allow for varied pruning ratios. \\
We then trained a LightGBM regressor \citep{giolin_lightgbm_2017} on the data to approximate the mapping from configuration to accuracy. On training, we achieved a train root mean squared error (RMSE) of 0.02282 and a test RMSE of 0.02902 with a $R^2$ value of 0.9863. After training, we extracted feature importance directly from the fitted LightGBM booster using its built-in feature importance mechanism. These scores were then normalized to yield comparable relative importances $\mathbf{I}$ as tabulated in Table \ref{tab:tab-3}. LightGBM was selected over linear regression models to capture the complex, non-linear interactions between distinct view-pruning ratios and their aggregate effect on classification accuracy.
\begin{table}[th]
	\centering
	\begin{adjustbox}{width=\textwidth}
		\begin{tabular}{|c|c|c|c|c|c|c|c|c|c|c|c|c|}
			\hline
			\textbf{View} 
			& 1 & 2 & 3 & 4 & 5 & 6 
			& 7 & 8 & 9 & 10 & 11 & 12 \\ 
			\hline
			\textbf{Importance (\%)} 
			& 7.2 & 10.5 & 7.9 & 7.7 & 7.8 & 8.6 
			& 9.1 & 8.7 & 8.6 & 7.6 & 8.3 & 7.9 \\ 
			\hline
		\end{tabular}
	\end{adjustbox}
	\caption{Importance of Each View}
	\label{tab:tab-3}
\end{table}
\subsubsection{Importance-aware pruning}
\label{sub:imp-aware}
Uniform pruning across views neglects the unequal contribution of individual views to the final classification accuracy. To address this, SlimEdge incorporates view-importance directly into the pruning allocation.\\
Let $\mathbf{I}=\left[I_1,\ldots,I_V\right]$ denote the normalized importance scores obtained in section \ref{sub:comp-view-imp}, where $I_v\in\left[0,1\right]$ and $\sum_{v} I_v=1.0$. Let $\mathbf{p^{\min}}=\left[p_1^{\min},\ldots,p_V^{\min}\right]$ be the pruning vector with the minimum pruning required to satisfy per-device memory constraints (Section \ref{sub:min-pruning-vec}). Any feasible pruning vector $p$ must satisfy: $p_v\geq p_v^{\min},\ \ \forall v$.\\
Additional pruning beyond the minimum is allocated inversely proportional to the view importance, such that higher-importance views are pruned less aggressively. Furthermore, we incorporate device performance into the pruning allocation. Let $D_v\geq0$ denote the performance factor of the device hosting view $v$. The Device Performance Factor $D_v$ serves as a normalized value for the computational throughput of each edge node. In our simulated environment, $D_v$ is used to scale the per-view inference time relative to a baseline throughput, such that lower values indicate reduced computational capacity and a higher likelihood of becoming a system bottleneck. By incorporating into the weighting term, SlimEdge ensures that hardware with lower computing capacity receives a higher pruning budget, thereby balancing the execution time across the heterogeneous cluster. Pruning pressure for each view is calculated using joint importance and device aware weighting term $W_v$, shown in Equation \ref{eq:eq-5}.
\begin{equation}
	W_v=\left(1-I_v\right)\left(1+D_v\right)
	\label{eq:eq-5}
\end{equation}
This weighting increases pruning for views that are both less informative and deployed on slower devices, while protecting high-importance views and views hosted on more capable hardware. During optimization, candidate pruning vectors $\mathbf{p}=\left[p_1,\ldots,p_V\right]$ are evaluated using $W_v$ biasing searches toward solutions that redistribute pruning capacity away from critical views and towards computationally constrained devices. Starting from the minimum pruning vector $\mathbf{p}^{\min}$ , additional pruning is distributed across views according to the normalized weights $W_v$. For a candidate solution with a total additional pruning budget $\Delta p_v^b$, the pruning ratio for each view is updated as in Equation \ref{eq:eq-6}, with $p_v$ clipped to the feasible interval $\left[0,1\right]$.
\begin{equation}
	p_v=p_v^{\min}+\Delta p_v^b\cdot\frac{W_v}{\sum_{u=1}^{V}W_u}
	\label{eq:eq-6}
\end{equation}

\subsubsection{Multi-objective formulation}
\label{multi-obj-form}
Optimization was guided by a composite objective that balances accuracy, model size, and inference time. The total Fitness score for a pruning configuration $p$ was defined as a weighted sum as shown in Equation \ref{eq:eq-7}.
\begin{equation}
	\mathcal{R}\left(\mathbf{p}\right)=\alpha \cdot \mathcal{R}_{acc}+\beta\cdot\mathcal{R}_{size}+\gamma\cdot\mathcal{R}_{time}
	\label{eq:eq-7}
\end{equation}
Here, $\mathcal{R}_{acc}$, $\mathcal{R}_{size}$ and $\mathcal{R}_{time}$ denote the accuracy, model size, and inference time fitness components, respectively, and $\alpha$, $\beta$ and $\gamma$ are their corresponding weighting coefficients.
\begin{enumerate}
	\item $\mathcal{R}_{acc}$ penalizes deviations below a user-defined accuracy threshold via an exponential decay (Equation \ref{eq:eq-8})
	\item $\mathcal{R}_{size}$ applies a Gaussian penalty to configurations exceeding the memory limit and assigns higher fitness to smaller feasible models (Equation \ref{eq:eq-9}).
	\item $\mathcal{R}_{time}$ similarly incentivizes lower latency (Equation \ref{eq:eq-10})
\end{enumerate}
The values of hyperparameters $\alpha$, $\beta$ and $\gamma$ are determined empirically as weights for the fitness score.
\begin{itemize}
	\item Accuracy Fitness Score $\mathcal{R}_{acc}$: This score quantifies the deviation of the configuration’s accuracy $\mathcal{A}(\mathbf{p})$ from the user-specified minimum $\mathcal{A}_{\min}$. It is defined as a piecewise exponential function of the difference $\Delta \mathcal{A}=\mathcal{A}\left(\mathbf{p}\right)-\mathcal{A}_{\min}$ (Equation \ref{eq:eq-8})
	\begin{equation}
		\mathcal{R}_{acc} =
		\begin{cases}
			\exp\!\left(-\dfrac{\Delta \mathcal{A}^{\kappa}}{2\,\sigma_r^2}\right),
			& 0 \le \Delta \mathcal{A} \le \epsilon \\[6pt]
			
			\exp\!\left(-\dfrac{\Delta \mathcal{A}}{2\,\sigma_r^2}\right),
			& \Delta \mathcal{A} > \epsilon \\[6pt]
			
			\exp\!\left(-\dfrac{\lvert \Delta \mathcal{A} \rvert^{\kappa}}{2\,\sigma_l^2}\right),
			& \text{otherwise}
		\end{cases}
		\label{eq:eq-8}
	\end{equation}
	Here, $\sigma_r$ and $\sigma_l$ adjust the curve slope for configurations above and below the threshold, respectively. The exponent $\kappa > 1$ controls the curvature of the penalty function and determines how aggressively deviations from the minimum accuracy are penalized. Larger values of $\kappa$ increase sensitivity to deviations near the threshold, while smaller values produce a smoother decay. In our implementation, $\kappa$ is selected empirically to balance constraint enforcement and optimization stability. Here, $\epsilon = 0.01$ defines a small tolerance region around the minimum accuracy requirement, allowing limited deviation without severe penalization.
	\item Model Size Fitness Score $\mathcal{R}_{size}$: To ensure hardware compatibility, $\mathcal{R}_{size}$ penalizes configurations exceeding memory limit $S_v^{\max}$ and rewards smaller footprints (Equation \ref{eq:eq-9}).
	\begin{equation}
		\mathcal{R}_{size} =
		\begin{cases}
			100\,\exp\!\left(-\dfrac{\lvert \Delta S \rvert^2}{2\,\sigma^2}\right),
			& \Delta S > 0 \\
			
			100 + \dfrac{S_v^{\max}}{S(p_v)},
			& \text{otherwise}
		\end{cases}
		\label{eq:eq-9}
	\end{equation}
	The score reaches its maximum when the model size of view $v$, $S_v\left(p_v\right)$ is minimized using a pruning ratio $p_v$, utilizing Gaussian decay to penalize violations $\Delta S=S_v\left(p_v\right)-{S_v}^{\max}$.
	\item Inference Time Fitness Score $\mathcal{R}_{time}$:
	This score incentivizes reduction of the end-to-end latency of the distributed framework by targeting the bottleneck subnetwork. Let \\ $\tilde{T}(\mathbf{p})=\frac{T^{\max}(\mathbf{p})}{T_{\text{baseline}}}$ denote the normalized maximum inference time across all active views, where $T_{\text{baseline}}$ is the latency of the unpruned MVCNN model. The inference-time fitness is defined as
	\begin{equation}
		\mathcal{R}_{time}
		=
		\exp\left(
		-\dfrac{\left(\tilde{T}(\mathbf{p})\right)^2}{2\sigma_t^2}
		\right)
		\label{eq:eq-10}
	\end{equation}
	where $\sigma_t$ controls the sensitivity of the fitness score to latency variations. Since $\tilde{T}(\mathbf{p})$ is dimensionless, this formulation ensures scale invariance and provides a smooth decay in fitness as latency increases relative to the baseline. Lower normalized latency yields a higher fitness score, promoting configurations that reduce the system bottleneck.
\end{itemize}

\subsubsection{Importance-aware sampling}
\label{sub:imp-aware-sampler}
To accelerate the convergence of the Non-Dominated Sorting Genetic Algorithm – II (NSGA-II) \citep{deb_fast_2002} (used in the section \ref{sub:optimization-algo}) in a high-dimensional pruning space, SlimEdge utilizes an importance-aware sampler. Rather than initializing the optimization with a purely random population, this sampler biases the initial search space using an importance-aware pruning vector, explained in section \ref{sub:imp-aware}. The sampler constructs the initial population $\mathbf{X}$ by injecting high-quality candidate solutions that respect both informational and hardware constraints. The initial population $\mathbf{X}$ consists of:
\begin{itemize}
	\item Importance-Aware Pruning Vector ($\mathbf{X}[0]$): This candidate represents a balanced solution where pruning intensity is distributed inversely proportional to view importance and computed from device performance factors (section \ref{sub:imp-aware}).
	\item Minimum Pruning Vector ($\mathbf{X}[1]$): This vector serves as the baseline feasible boundary, ensuring the model fits within the physical memory of each edge device as determined in section \ref{sub:min-pruning-vec}.
	\item Beta-Distributed Samples ($\mathbf{X}\left[2\ldots n\right]$) The remainder of the population is generated using a Beta distribution to explore the feasible region around the initial vectors. The Beta distribution was selected because its bounded support on $[0,1]$ matches the feasible pruning domain, and its shape parameters allow importance-aware skewing of samples toward higher pruning ratios for low-importance views while preserving conservative sampling for critical views.
\end{itemize}
For each view $v$, the shape parameters of the Beta distribution are based on its normalized importance $I_v$. The shape parameter $\alpha$ is calculated as shown in Equation \ref{eq:eq-11} where the constants were selected arbitrarily based on the performance. 
\begin{equation}
	\alpha_v=\left(1-I_v\right)\cdot \zeta
	\label{eq:eq-11}
\end{equation} 
where $\zeta > 0$ is a scaling hyperparameter controlling the skewness of the distribution. Larger values of $\zeta$ increase the concentration of samples toward higher pruning ratios for views with low importance, while smaller values result in a more uniform exploration of the feasible space. In this work, $\zeta$ is selected empirically to balance initialization bias and population diversity.
The second shape parameter is fixed at $\beta = 1.0$, yielding a $\text{Beta}(\alpha_v, 1)$ distribution. When $I_v$ is small, $\alpha_v$ becomes large, producing a distribution skewed toward 1 and thus favoring aggressive pruning. Conversely, when $I_v$ is large, $\alpha_v$ approaches zero, shifting the distribution toward lower pruning ratios and thereby protecting highly informative views.
All generated samples are clipped to the feasible range $\left[p_v^{\min}, 1.0\right]$, ensuring that every candidate solution in the initial population satisfies local memory constraints of active hardware. For any device identified as offline, the sampler fixes $p_v = 1.0$, effectively pruning the entire subnetwork to maintain robustness under device failure.

\subsubsection{Optimization algorithm: NSGA-II}
\label{sub:optimization-algo}
To solve the multi-objective pruning problem, SlimEdge employs the Non-Dominated Sorting Genetic Algorithm II (NSGA-II) \citep{deb_fast_2002} to explore the non-convex Pareto front defined by latency, accuracy deviation, and a composite reward metric. NSGA-II is selected due to the non-convex, high-dimensional search space and the presence of competing objectives, where gradient-based or single-objective methods are insufficient. The algorithm is initialized with parameters derived from both the application requirements and the physical edge environment.
\begin{itemize}
	\item Output Vector - Pruning variables: A vector $\textbf{p}=\left[p_1,p_2,\ldots,p_{12}\right]$ representing the pruning ratios for the 12 view-specific feature extractors.
	\item Device Status: An online mask, $DevOnline_v=0$ for failed or offline nodes, forcing $p_v=1.0$.
	\item Hardware Metrics: Per-device memory capacities ($M_v$) and normalized computing performance factors ($D_v$).
\end{itemize}
 Three objective functions were defined, and the algorithm simultaneously minimizes all three objectives to explore trade-offs between latency, accuracy deviation, and composite reward. All objectives are normalized to ensure scale invariance during evolutionary optimization.
\begin{enumerate}
	\item Latency Minimization $f_1$ (Equation \ref{eq:eq-12}):
	\begin{equation}
		f_1\left(\mathbf{p}\right)=\frac{{\max}_{v\in V_{online}}{\left(T_v\left(p_v,D_v\right)\right)}}{T_{baseline}}
		\label{eq:eq-12}
	\end{equation}
	Where $T_v$ is the inference time of the view $v$ and $T_{baseline}$ is the unpruned model’s latency. This ensures the slowest device does not bottleneck the distributed system.
	\item Accuracy Deviation Minimization $f_2$ (Equation \ref{eq:eq-13}):
	\begin{equation}
		f_2\left(\mathbf{p}\right)=\left|\mathcal{A}\left(\mathbf{p}\right)-\mathcal{A}_{\min}\right|
		\label{eq:eq-13}
	\end{equation}
	Where $\mathcal{A}(\mathbf{p})$ is the predicted multi-view accuracy. This objective minimizes deviation from the target accuracy and prioritizes solutions that satisfy or slightly exceed $\mathcal{A}_{\min}$. Among feasible solutions satisfying $A(\textbf{p}) \ge A_{\min}$, minimizing deviation discourages excessive overfitting to accuracy at the expense of latency and model size.
	\item Reward Maximization $f_3$ (Equation \ref{eq:eq-14}):
	\begin{equation}
		f_3(\mathbf{p}) = -\frac{\mathcal{R}(\mathbf{p})}{\mathcal{R}_{\max}}
		\label{eq:eq-14}
	\end{equation}
	The composite reward $\mathcal{R}(\mathbf{p})$, defined in Section \ref{multi-obj-form}, is incorporated as a scalar objective. $R_{\max}$ denotes the maximum reward observed in the current population, ensuring bounded normalization within $[0,1]$. Since NSGA-II performs minimization, the negative normalized reward is minimized. This objective promotes globally balanced configurations by coupling accuracy, memory efficiency, and latency into a single scalar metric.
\end{enumerate}
While latency and accuracy are optimized explicitly, the reward objective introduces a coupled system-level trade-off, guiding the search toward solutions that jointly satisfy performance and resource-efficiency criteria. Furthermore, the search space is strictly bound by two primary hard constraints:
\begin{enumerate}
	\item Accuracy Constraint: $\mathcal{A}\left(\mathbf{p}\right)\geq\mathcal{A}_{\min}$
	\item Memory Constraint: $\forall v,S_v\left(p_v\right)\le M_v$, where $S_v$ is the model size at pruning ratio $p_v$. \\
	The algorithm uses the minimum pruning vector $\mathbf{p}^{\min}$  as the lower bound for variables to satisfy this constraint automatically.
\end{enumerate}
The algorithm executes through $N_{gen}=200$ generations with a population size of 400 to converge on a set of optimal pruning ratios. To accelerate convergence, the initial population is seeded by the Importance-Aware Sampler (Section \ref{sub:imp-aware-sampler}) which injects the minimum pruning vector and importance-aware pruning vector into the first generation. In each generation, candidate pruning vectors are evaluated against three objective functions $(f_1, f_2, f_3)$. Solutions are then ranked using non-dominated sorting and crowding distance \citep{deb_fast_2002} to maintain a diverse Pareto front that explores multiple trade-offs between speed and accuracy. Feasibility is enforced by checking each candidate against accuracy and memory constraints. Configurations that fail to meet these thresholds are penalized or discarded during the selection process. \\
Device availability is incorporated into the optimization process. If a device is marked offline, its pruning ratio is fixed to 1.0 and pruning is redistributed among the remaining active views. After convergence, the best feasible solution is selected and deployed to the corresponding edge devices.

\section{Experimental setup}
\label{experimental-setup}
Extensive testing was conducted across diverse configurations to evaluate the performance of the proposed algorithm under heterogeneous and constrained deployment conditions.
\subsection{Device simulation}
To evaluate SlimEdge, we developed a parametric device simulation framework that models heterogeneous edge node behavior. Each simulated device is assigned a unique memory capacity $M_v$ and a performance factor $D_v$ sampled from uniform distributions between $[80, 300]$ and $[0, 1]$ respectively to replicate a mixed-tier hardware ecosystem. Rather than measuring wall-clock time on specific physical boards, latency is quantified via a normalized inference time model, where $D_v$ scales the base inference cost of the feature extractor. This allows for the testing of SlimEdge logic across a broader spectrum of hardware configurations than a static physical cluster would permit. A 30\% device failure probability was chosen to emulate volatile distributed edge environments with intermittent node availability. This stress-testing configuration exceeds typical deployment conditions and enables systematic evaluation of robustness under severe failure scenarios. The minimum required accuracy $\mathcal{A}_{\min}$ was sampled by first drawing $u \sim \text{Beta}(4.0, 1.2)$, and then linearly mapping it to the interval $[0.65, 0.88]$ as $A_{\min} = 0.65 + 0.23\,u$. This sampling strategy biases the experiments toward higher accuracy thresholds while still allowing moderate variability.
\subsection{Model and dataset}
A custom VGG11 model was employed, pre-trained on the ImageNet1K dataset \citep{deng_imagenet_2009}, which was then fine-tuned on the ModelNet40 dataset \citep{wu_3d_2015}. This dataset includes 12 azimuthal viewpoints per object, each spaced 30 degrees apart, at a fixed elevation of 30 degrees. Furthermore, the final classification layer of the VGG11 architecture was modified from 1000 output units to 40 output units to match the number of classes of the ModelNet40 dataset. Upon training, the model achieved a testing accuracy of 88.30\%.
\subsection{Evaluation metrics}
SlimEdge was evaluated on the following metrics:
\begin{enumerate}
	\item Accuracy: Classification performance of the framework after optimization
	\item Inference time reduction: The decrease in modeled latency of the distributed system compared to a non-optimized baseline.
\end{enumerate}

\section{Results}
\label{results}
SlimEdge was evaluated across 1,000 simulated configurations to assess its robustness under heterogeneous device constraints. The optimization framework successfully identified feasible solutions in 95.30\% of the cases. In the remaining 4.70\% of configurations, no feasible solution satisfied the minimum accuracy constraint $\mathcal{A}_{\min}$. These failures occurred when the imposed memory, latency, and device-availability constraints rendered the optimization problem infeasible. The key experiments and corresponding findings are presented below.
\begin{figure}[ht]
	\centering
	\includegraphics[width=0.8\textwidth]{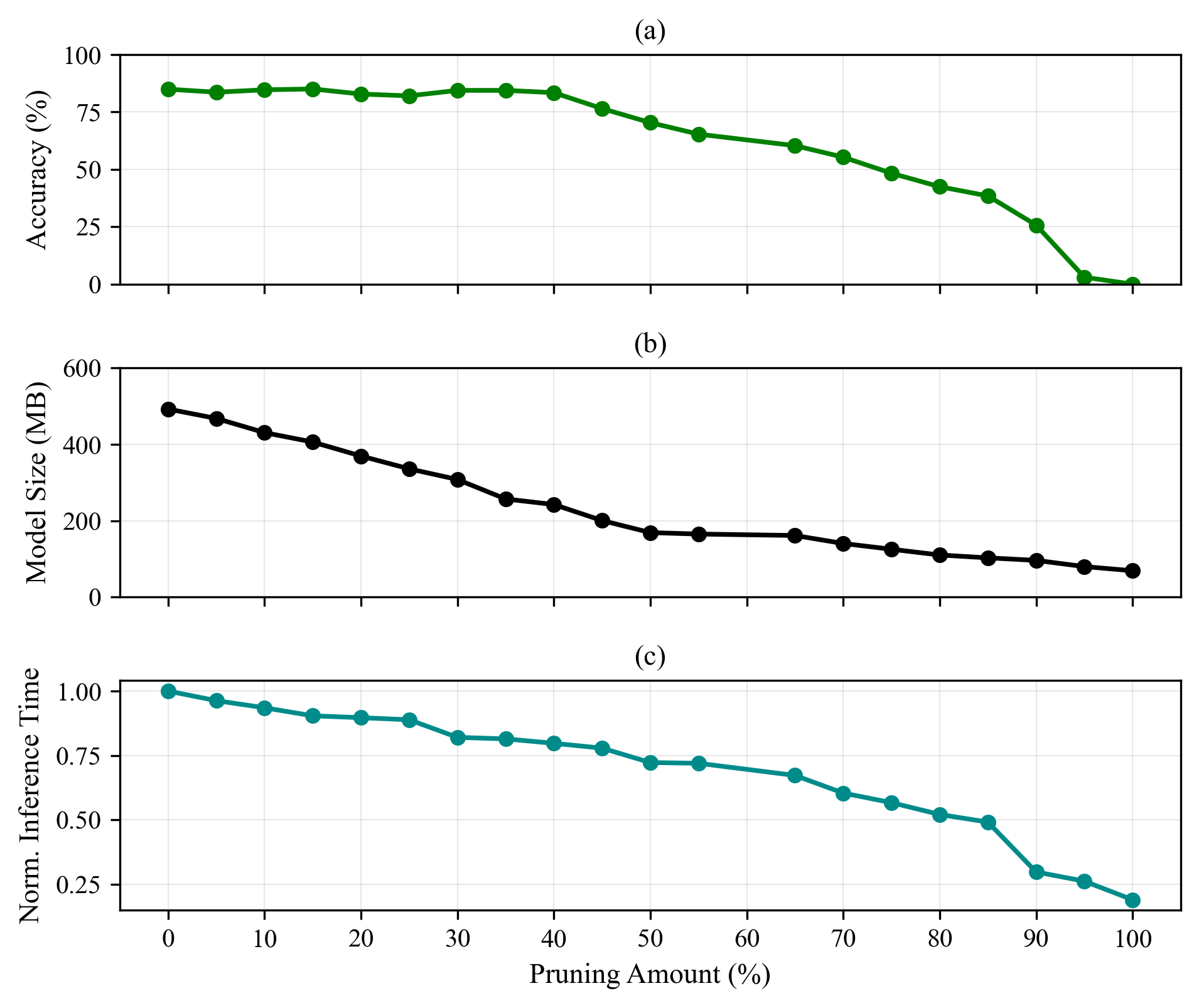}
	\caption{Effect of pruning on the characteristics of the MVCNN model}\label{fig:fig-6}
\end{figure}
\subsection{Impact of pruning on accuracy}
All views of the MVCNN model were pruned uniformly using filter pruning and subsequently fine-tuned on the ModelNet40 dataset \citep{wu_3d_2015} until validation accuracy converged. Fine-tuning was performed to recover performance degradation resulting from the reduced representational capacity caused by filter removal. The resulting mean class accuracy after pruning is shown in Figure \ref{fig:fig-6}a. The combined effect of filter pruning and fine-tuning allows the model to regain accuracy, even at high pruning ratios. As only the least essential filters are removed at each step, the model can retain the filters that contribute most to feature representation ensuring minimal degradation in performance. Fine-tuning after each pruning step further adapts the remaining parameters to compensate for the reduced performance, preserving accuracy. The superiority of SlimEdge over uniform pruning baselines is particularly evident at high compression ratios. While uniform pruning treats all views as equal, SlimEdge protects critical perspectives identified in Table 1. By redistributing the pruning budget away from high-importance views and toward less informative or slower simulated nodes, the framework maintains accuracy thresholds that uniform methods struggle to achieve.
\subsection{Effect of pruning on model size}
On pruning all views uniformly, the model size of the MVCNN model decreases with an increase in pruning. This is due to the number of filters being removed, where each element in each filter is 4 bytes in a float32 model. The model size decreases in a near linear order, as shown in Figure \ref{fig:fig-6}b. This enables us to fit the pruned models in resource-constrained devices, effectively satisfying resource constraints.
\subsection{Effect of pruning on inference time}
Similar to the effect of pruning on model size, a near linear decrease in inference time was observed. The inference time was normalized using min-max normalization and has been plotted in Figure \ref{fig:fig-6}c. The inference time of the MVCNN model was significantly reduced by pruning the model. The reduction in the number of filters after pruning reduces the number of computations required, thus speeding up inference. Beyond simple filter reduction, the inclusion of the device performance factor in the optimization loop ensures that simulated nodes with higher computational costs are pruned more aggressively. This directly addresses the “bottleneck subnetwork” problem, where the slowest device determines the end-to-end latency of the distributed system.
\subsection{Ablation study}
\begin{table}[ht]
	\centering
	\begin{adjustbox}{width=\textwidth}
		\begin{tabular}{|p{2.3cm}|p{4cm}|p{3.1cm}|p{1.5cm}|p{1.4cm}|}
			\hline
			\textbf{Configuration} & \textbf{Pruning Logic} & \textbf{Key Parameters} & \textbf{Accuracy} & \textbf{Speedup}\textsuperscript{*} \\ 
			\hline
			
			Baseline 1 & Uniform: Fixed pruning ratio for all views & $p_v = 0.85$ & 32.15\% & 4.88$\times$ \\ 
			\hline
			
			Baseline 2 & Hardware-Aware: Pruning to fit memory limits & $M_v \sim \mathcal{U}(80, 300)\,\text{MB}$ & 59.96\% & 4.06$\times$ \\ 
			\hline
			
			SlimEdge & Full Joint Optimization: View, Device and Failure Optimization & $p_v, M_v, I_v, D_v$ & 80.00\% & 5.47$\times$ \\ 
			\hline
			
		\end{tabular}
	\end{adjustbox}
	\footnotesize{\textsuperscript{*}Speedup measured relative to the unpruned MVCNN model}
	\caption{Incremental Ablation of SlimEdge Optimization Logic}
	\label{tab:tab-4}
\end{table}
The ablation study shown in Table \ref{tab:tab-4} demonstrates the cumulative benefit of integrating system-level constraints into the pruning process. While Baseline 1 provides a basic metric for model degradation under uniform pruning, Baseline 2 shows that simply enforcing memory constraints $M_v$ using the Newton-Raphson method improves the speedup to 4.06$\times$ by tailoring model sizes to individual simulated nodes. Finally, the full SlimEdge configuration achieves the highest accuracy and a 5.47$\times$ speedup by incorporating view importance $I_v$ and device performance $D_v$ into the optimization loop. This confirms that the framework prevents bottlenecking by pruning slower simulated devices more aggressively while protecting critical informational views.
\subsection{Experiments}
Experiments were conducted on SlimEdge to validate its performance under various parameters and constraints. The framework showed robust performance in varied conditions of device-failure and device configurations and was able to provide a model tailored for each edge device.\\
For every experiment conducted, a modified VGG11 base model was used as the MVCNN model \citep{su_multi-view_2015} to train on the ModelNet40 dataset. This model had a mean class accuracy of 88.30\% and a model size of 506.8 megabytes (MB). All reported speedups are measured relative to the unpruned MVCNN baseline. Speedups reported for heterogeneous and bottleneck-focused experiments are additionally compared against unoptimized models to highlight the benefit of device-aware and importance-aware allocation.
\subsubsection{Experiment 1: simulating performance under optimal conditions}
\begin{figure}[ht]
	\centering
	\includegraphics[width=1.0\textwidth]{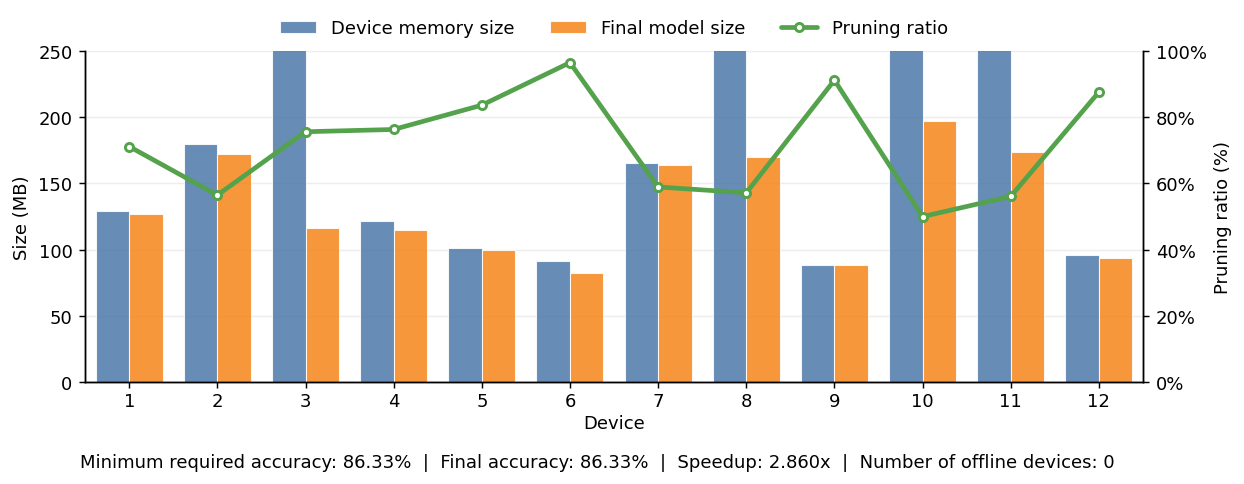}
	\caption{Optimized pruning configuration under full device availability showing that SlimEdge satisfies the minimum accuracy requirement (86.33\%) while achieving a 2.86$\times$ latency speedup within all device memory constraints.}
    \label{fig:fig-7}
\end{figure}

This experiment evaluates the performance of SlimEdge under optimal conditions where all devices are online and functioning. For this experiment, we defined the minimum required accuracy $\mathcal{A}_{\min}$ of 86.33\%, with all 12 devices online. The per-device information for this experiment is detailed in Table \ref{tab:tab-5-exp-1}.\\
\begin{table}[ht]
	\centering
	\begin{adjustbox}{width=\textwidth}
		\begin{tabular}{|p{2cm}|c|c|c|c|c|c|c|c|c|c|c|c|}
			\hline
			\textbf{Device} 
			& 1 & 2 & 3 & 4 & 5 & 6 & 7 & 8 & 9 & 10 & 11 & 12 \\ 
			\hline
			
			\textbf{Device Memory Capacity $M_v$ (MB)} 
			& 129.40 & 180.06 & 294.42 & 122.05 & 101.23 & 91.83 
			& 165.59 & 278.59 & 88.50 & 269.13 & 273.49 & 96.24 \\ 
			\hline
			
			\textbf{Device Performance Factor $D_v$} 
			& 0.033 & 0.101 & 0.078 & 0.049 & 0.117 & 0.122 
			& 0.049 & 0.111 & 0.046 & 0.106 & 0.052 & 0.130 \\ 
			\hline
		\end{tabular}
	\end{adjustbox}
	\caption{Per-device memory capacities and performance factors under full device availability used to evaluate SlimEdge feasibility and speedup under optimal operating conditions.}
	\label{tab:tab-5-exp-1}
\end{table}
Upon having SlimEdge optimize the system, we observe the following results:
\begin{enumerate}
	\item The accuracy of the optimized system meets the minimum accuracy requirement $\mathcal{A}_{\min}$, providing a final accuracy of 86.33\%.
	\item A speedup in inference time of 2.86$\times$ is observed.
	\item All the pruned models fit within the memory of the device associated with that view.
\end{enumerate}
These results have been plotted in Figure \ref{fig:fig-7}.
\subsubsection{Experiment 2: simulating performance under 33\% device failure}
\begin{figure}[ht]
 	\centering
 	\includegraphics[width=1.0\textwidth]{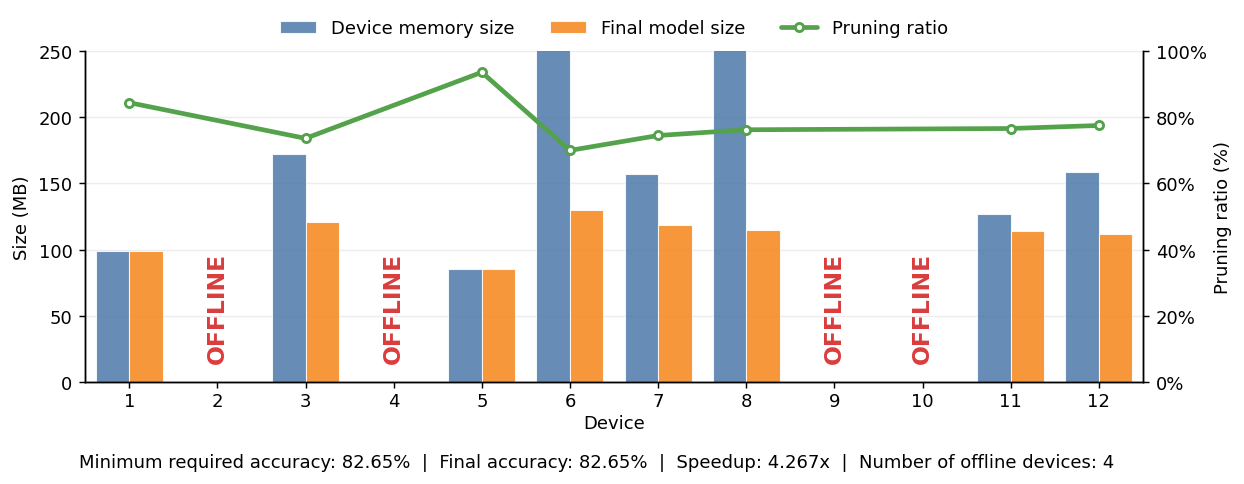}
 	\caption{Optimized pruning configuration under 33\% device failure demonstrating that SlimEdge maintains the target accuracy (82.65\%) and achieves a 4.26$\times$ speedup by redistributing pruning across the remaining active devices.}
    \label{fig:fig-8}
 \end{figure}

To validate the performance of SlimEdge under constrained conditions, we tested SlimEdge under tight physical constraints; while simulating device failure, we conducted experiment 2. Here, we defined the minimum required accuracy $\mathcal{A}_{\min}$ to be 82.65\%. Furthermore, we set up 4 devices, namely, device 2, 4, 9 and 10 to be offline (non-functioning). The per-device information has been tabulated below in Table \ref{tab:tab-6-exp-2}.
\begin{table}[ht]
	\centering
	\begin{adjustbox}{width=\textwidth}
		\begin{tabular}{|p{2cm}|c|c|c|c|c|c|c|c|c|c|c|c|}
			\hline
			\textbf{Device} 
			& 1 & 2 & 3 & 4 & 5 & 6 & 7 & 8 & 9 & 10 & 11 & 12 \\ 
			\hline
			
			\textbf{Device Memory Capacity $M_v$ (MB)} 
			& 99.11 & N/A & 172.21 & N/A & 85.50 & 268.88 
			& 157.09 & 259.93 & N/A & N/A & 127.11 & 159.01 \\ 
			\hline
			
			\textbf{Device Performance Factor $D_v$} 
			& 0.029 & N/A & 0.130 & N/A & 0.078 & 0.112 
			& 0.945 & 0.040 & N/A & N/A & 0.094 & 0.080 \\ 
			\hline
		\end{tabular}
	\end{adjustbox}
	\caption{Per-device configuration under 33\% device failure, showing heterogeneous memory budgets and performance factors for the remaining active devices.}
	\label{tab:tab-6-exp-2}
\end{table}
SlimEdge was made to optimize this setup, where it was able to optimize the setup successfully. The key findings of this experiment are detailed below:
\begin{enumerate}
	\item 	An accuracy of 82.65\% was achieved which matched the minimum required accuracy $\mathcal{A}_{\min}$.
	\item Inference time was sped up by a factor of 4.26$\times$.
	\item The size of the model at each device was less than the device memory, thus allowing the model to fit in memory
\end{enumerate}
 In this 33\% failure scenario, SlimEdge maintained the target accuracy of 82.65\% by reallocating the informational load to the 8 remaining active views. The distribution as well as the memory capacities and model sizes have been plotted in Figure \ref{fig:fig-8}.
 \subsubsection{Experiment 3: simulating performance under 50\% device failure}
 SlimEdge has shown its efficacy even when 4 of 12 devices fail. Now to test the limitations of the algorithm, an even more intensive experiment was conducted. Here, 50\% of the devices fail, i.e., 6 of 12 devices are offline. Devices 1, 2, 4, 7, 9, 12 were set to be offline. A minimum accuracy requirement $\mathcal{A}_{\min}$ was defined as 75.22\%, which would push the boundaries of the algorithm. The per-view information has been tabulated in Table \ref{tab:tab-7-exp-3}.
 \begin{table}[ht]
 	\centering
 	\begin{adjustbox}{width=\textwidth}
 		\begin{tabular}{|p{2cm}|c|c|c|c|c|c|c|c|c|c|c|c|}
 			\hline
 			\textbf{Device} 
 			& 1 & 2 & 3 & 4 & 5 & 6 & 7 & 8 & 9 & 10 & 11 & 12 \\ 
 			\hline
 			
 			\textbf{Device Memory Capacity $M_v$ (MB)} 
 			& N/A & N/A & 156.23 & N/A & 219.92 & 269.83 
 			& N/A & 261.22 & N/A & 131.13 & 293.40 & N/A \\ 
 			\hline
 			
 			\textbf{Device Performance Factor $D_v$} 
 			& N/A & N/A & 0.150 & N/A & 0.049 & 0.049 
 			& N/A & 0.082 & N/A & 0.058 & 0.036 & N/A \\ 
 			\hline
 		\end{tabular}
 	\end{adjustbox}
 	\caption{Per-device configuration under 50\% device failure, highlighting severe resource heterogeneity and reduced device availability used to stress-test SlimEdge robustness.}
 	\label{tab:tab-7-exp-3}
 \end{table}
 Having SlimEdge optimized the setup, the following were observed:
 \begin{enumerate}
 	\item The minimum accuracy requirement $\mathcal{A}_{\min}$ of 75.22\% was met after optimization.
 	\item A speedup in inference time by a factor of 4.70$\times$ was observed.
 	\item The size of the model at each device was smaller than the device's memory, which enabled the model to fit within the memory of the edge device.
 \end{enumerate}
  \begin{figure}[ht]
 	\centering
 	\includegraphics[width=1.0\textwidth]{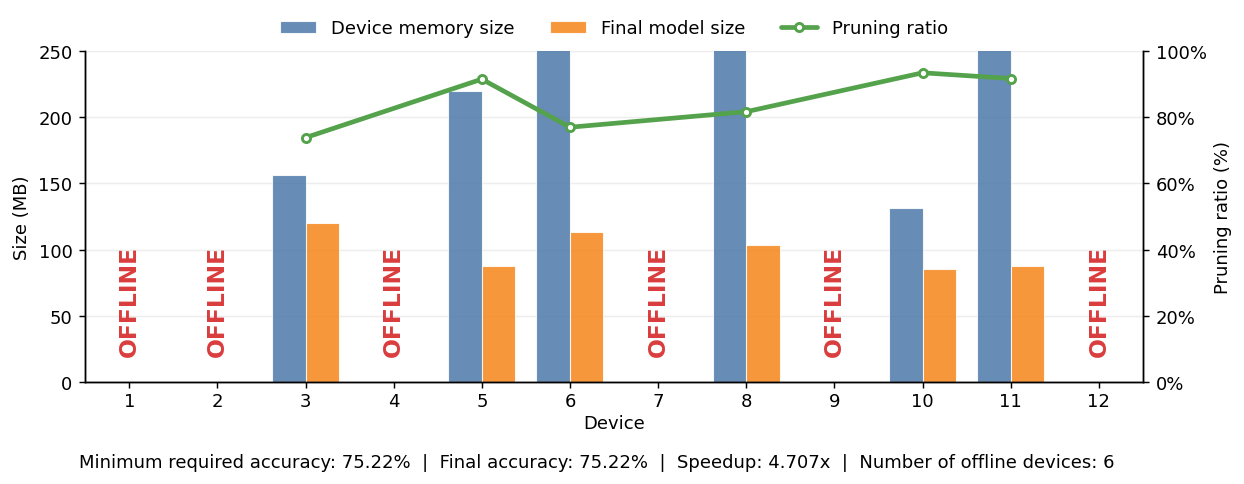}
 	\caption{Optimized pruning configuration under 50\% device failure showing that SlimEdge preserves the minimum accuracy (75.22\%) and achieves a 4.70$\times$ speedup despite severe device unavailability.}
    \label{fig:fig-9}
 \end{figure}
 In this 50\% failure scenario, SlimEdge maintained the target accuracy of 75.22\% by reallocating the informational load to the 6 remaining active views, a scenario where static or uniform pruning strategies typically fail to satisfy accuracy constraints. The distribution of pruning across the functioning devices is shown in Figure \ref{fig:fig-9}. The ability of SlimEdge to optimize such constrained environments shows the efficacy of SlimEdge and its potential use in distributed edge computing.

\section{Conclusion}
\label{conclusion}
SlimEdge presents a framework for dynamic pruning techniques for distributed inference systems, coupled with fault-tolerance logic. Through rigorous experimentation, the effectiveness of SlimEdge in optimizing model performance while striking a balance between accuracy and efficiency by matching minimum accuracies $\mathcal{A}_{\min}$ of 86.33\%, 82.65\% and 75.22\% with speedups of up to 4.70$\times$ has been demonstrated. Furthermore, the algorithm was able to make these optimizations while there were multiple device failures. The integration of the NSGA-II \citep{deb_fast_2002} algorithm along with the objective functions to optimize the model has yielded significant improvements in the deployment and optimization of the MVCNN \citep{su_multi-view_2015} and other similar frameworks. The implications of such optimizations for edge devices have the potential to reshape the deployment of such models on resource-constrained devices.\\
Across all experiments, the framework consistently identified feasible allocations that met accuracy and size requirements, while minimizing end-to-end inference latency under heterogeneous hardware constraints. The results demonstrate that explicitly modeling view importance and device capability enables SlimEdge to avoid the inefficiencies of uniform compression, particularly in failure-prone distributed settings. By reallocating pruning budgets away from high-importance views and toward less informative or slower devices, the framework preserves global performance while respecting strict per-device memory limits. More broadly, SlimEdge illustrates that distributed inference systems benefit from treating compression as a system-level optimization problem rather than a static, model-centric operation. The combination of importance-aware pruning, constrained multi-objective optimization, and failure-aware deployment allows the framework to adapt dynamically to changing device availability without retraining the base model. This decoupling of deployment-time optimization from model training is particularly advantageous in edge environments, where hardware heterogeneity and intermittent failures are the norm rather than the exception. While this work focuses on MVCNN-based 3D object recognition, the proposed formulation is not limited to multi-view vision models. The underlying principles of importance estimation, constraint-aware pruning, and Pareto-optimal allocation naturally extend to other distributed neural architectures, including sensor-fusion systems and multi-stream perception pipelines. Future work may explore tighter integration with online performance monitoring, alternative importance estimation techniques, and extension to quantized or mixed-precision models to further reduce deployment overhead.\\
In summary, SlimEdge provides a practical and extensible framework for deploying deep, distributed neural networks on constrained edge hardware. By jointly accounting for informational asymmetry, device heterogeneity, and failure resilience, it offers a principled pathway toward scalable and reliable edge intelligence in real-world environments.

\bibliographystyle{elsarticle-harv} 
\bibliography{SlimEdge-v2}

@misc{su_multi-view_2015,
  title = {Multi-view Convolutional Neural Networks for 3D Shape Recognition},
  author = {Su, Hang and Maji, Subhransu and Kalogerakis, Evangelos and Learned-Miller, Erik},
  year = {2015},
  month = {sep},
  eprint = {1505.00880},
  archivePrefix = {arXiv},
  primaryClass = {cs.CV},
  doi = {10.48550/arXiv.1505.00880}
}

@article{angrish_mvcnn_2021,
  title = {MVCNN++: Computer-Aided Design Model Shape Classification and Retrieval Using Multi-View Convolutional Neural Networks},
  author = {Angrish, Atin and Bharadwaj, Akshay and Starly, Binil},
  journal = {Journal of Computing and Information Science in Engineering},
  volume = {21},
  number = {1},
  pages = {011001},
  year = {2021},
  month = {feb},
  doi = {10.1115/1.4047486}
}

@article{yu_latent-mvcnn_2020,
  title = {Latent-MVCNN: 3D Shape Recognition Using Multiple Views from Pre-defined or Random Viewpoints},
  author = {Yu, Qian and Yang, Chengzhuan and Fan, Honghui and Wei, Hui},
  journal = {Neural Processing Letters},
  volume = {52},
  number = {1},
  pages = {581--602},
  year = {2020},
  month = {aug},
  doi = {10.1007/s11063-020-10268-x}
}

@article{liang_pruning_2021,
  title = {Pruning and Quantization for Deep Neural Network Acceleration: A Survey},
  author = {Liang, Tailin and Glossner, John and Wang, Lei and Shi, Shaobo and Zhang, Xiaotong},
  journal = {Neurocomputing},
  volume = {461},
  pages = {370--403},
  year = {2021},
  month = {oct},
  doi = {10.1016/j.neucom.2021.07.045}
}

@misc{baker_designing_2017,
  title = {Designing Neural Network Architectures using Reinforcement Learning},
  author = {Baker, Bowen and Gupta, Otkrist and Naik, Nikhil and Raskar, Ramesh},
  year = {2017},
  month = {mar},
  eprint = {1611.02167},
  archivePrefix = {arXiv},
  primaryClass = {cs.LG},
  doi = {10.48550/arXiv.1611.02167}
}

@misc{chen_searching_2018,
  title = {Searching for Efficient Multi-Scale Architectures for Dense Image Prediction},
  author = {Chen, Liang-Chieh and Collins, Maxwell D. and Zhu, Yukun and Papandreou, George and Zoph, Barret and Schroff, Florian and Adam, Hartwig and Shlens, Jonathon},
  year = {2018},
  month = {sep},
  eprint = {1809.04184},
  archivePrefix = {arXiv},
  primaryClass = {cs.CV},
  doi = {10.48550/arXiv.1809.04184}
}

@article{zhao_deepthings_2018,
  title = {DeepThings: Distributed Adaptive Deep Learning Inference on Resource-Constrained IoT Edge Clusters},
  author = {Zhao, Zhuoran and Barijough, Kamyar Mirzazad and Gerstlauer, Andreas},
  journal = {IEEE Transactions on Computer-Aided Design of Integrated Circuits and Systems},
  volume = {37},
  number = {11},
  pages = {2348--2359},
  year = {2018},
  month = {nov},
  doi = {10.1109/TCAD.2018.2858384}
}

@article{deb_fast_2002,
  title = {A Fast and Elitist Multiobjective Genetic Algorithm: NSGA-II},
  author = {Deb, K. and Pratap, A. and Agarwal, S. and Meyarivan, T.},
  journal = {IEEE Transactions on Evolutionary Computation},
  volume = {6},
  number = {2},
  pages = {182--197},
  year = {2002},
  month = {apr},
  doi = {10.1109/4235.996017}
}

@misc{molchanov_pruning_2017,
  title = {Pruning Convolutional Neural Networks for Resource Efficient Inference},
  author = {Molchanov, Pavlo and Tyree, Stephen and Karras, Tero and Aila, Timo and Kautz, Jan},
  year = {2017},
  month = {jun},
  eprint = {1611.06440},
  archivePrefix = {arXiv},
  primaryClass = {cs.LG},
  doi = {10.48550/arXiv.1611.06440}
}

@misc{frankle_lottery_2019,
  title = {The Lottery Ticket Hypothesis: Finding Sparse, Trainable Neural Networks},
  author = {Frankle, Jonathan and Carbin, Michael},
  year = {2019},
  month = {mar},
  eprint = {1803.03635},
  archivePrefix = {arXiv},
  primaryClass = {cs.LG},
  doi = {10.48550/arXiv.1803.03635}
}

@inproceedings{yao_efficient_2019,
  title = {Efficient Implementation of Convolutional Neural Networks with End to End Integer-Only Dataflow},
  author = {Yao, Yiwu and Dong, Bin and Li, Yuke and Yang, Weiqiang and Zhu, Haoqi},
  booktitle = {2019 IEEE International Conference on Multimedia and Expo (ICME)},
  pages = {1780--1785},
  year = {2019},
  month = {jul},
  doi = {10.1109/ICME.2019.00306}
}

@misc{wu_3d_2015,
  title = {3D ShapeNets: A Deep Representation for Volumetric Shapes},
  author = {Wu, Zhirong and Song, Shuran and Khosla, Aditya and Yu, Fisher and Zhang, Linguang and Tang, Xiaoou and Xiao, Jianxiong},
  year = {2015},
  month = {apr},
  eprint = {1406.5670},
  archivePrefix = {arXiv},
  primaryClass = {cs.CV},
  doi = {10.48550/arXiv.1406.5670}
}

@inproceedings{giolin_lightgbm_2017,
  title = {LightGBM: A Highly Efficient Gradient Boosting Decision Tree},
  author = {Ke, Guolin and Qi, Meng and Finley, Thomas and Wang, Taifeng and Chen, Wei and Ma, Weidong and Ye, Qiwei and Liu, Tie-Yan},
  booktitle = {Advances in Neural Information Processing Systems 30 (NIPS 2017)},
  pages = {3149--3157},
  year = {2017}
}

@inproceedings{deng_imagenet_2009,
  title = {ImageNet: A Large-Scale Hierarchical Image Database},
  author = {Deng, Jia and Dong, Wei and Socher, Richard and Li, Li-Jia and Li, Kai and Fei-Fei, Li},
  booktitle = {2009 IEEE Conference on Computer Vision and Pattern Recognition},
  pages = {248--255},
  year = {2009},
  month = {jun},
  doi = {10.1109/CVPR.2009.5206848}
}

@article{mckay_comparison_1979,
  title = {A Comparison of Three Methods for Selecting Values of Input Variables in the Analysis of Output from a Computer Code},
  author = {McKay, M. D. and Beckman, R. J. and Conover, W. J.},
  journal = {Technometrics},
  volume = {21},
  number = {2},
  pages = {239--245},
  year = {1979},
  month = {may},
  doi = {10.2307/1268522}
}

@book{raphson_analysis_1697,
  title = {Analysis Aequationum Universalis},
  author = {Raphson, Joseph},
  year = {1697},
  publisher = {Tho. Braddyll},
  doi = {10.3931/E-RARA-13516},
}

@article{das_toward_2024,
  title = {Toward Energy-Efficient Collaborative Inference Using Multisystem Approximations},
  author = {Das, Arghadip and Ghosh, Soumendu Kumar and Raha, Arnab and Raghunathan, Vijay},
  journal = {IEEE Internet of Things Journal},
  volume = {11},
  number = {10},
  pages = {17989--18004},
  year = {2024},
  month = {may},
  doi = {10.1109/JIOT.2024.3365306}
}

\end{document}